\documentclass[preprint2]{emulateapj}
\usepackage{graphicx, amsmath}
\usepackage{bm, color}
\usepackage{booktabs}
\usepackage{multirow}
\graphicspath{{./fig/}{./png/}}

\definecolor{brown}{rgb}{0.42,0.24,0.07}
\definecolor{darkgreen}{rgb}{0.0,0.6,0.00}
 \definecolor{purple}{rgb}{0.7,0.0,0.7}

\newcommand{\op}{\:\!}
\newcommand{\tp}{\;\!}
\newcommand{\Mes}{\emph{MESSENGER\ }}

\newcommand{\EQ}{\begin{equation}}
\newcommand{\EN}{\end{equation}}
\newcommand{\vv}{\mbox{\boldmath $v$} {}}
\newcommand{\uu}{\mbox{\boldmath $u$} {}}
\newcommand{\Sec}[1]{Section~\ref{#1}}
\newcommand{\Eq}[1]{Equation~(\ref{#1})}
\newcommand{\Fig}[1]{Figure~\ref{#1}}

\date{\today,~ $ $Revision: 1.1 $ $}
\begin{document}

\title{Explaining Mercury's Density through Magnetic Erosion}
\author{Alexander Hubbard\altaffilmark{1}}
\altaffiltext{1}{Department of Astrophysics, American Museum of Natural History, New York, NY 10024-5192, USA}
\email{ahubbard@amnh.org}

\begin{abstract}
In protoplanetary disks, dust grains rich in metallic iron can attract each other magnetically.  If they are
magnetized to values near saturation, the magnetically induced collision speeds are
high enough to knock off the non-magnetized, loosely bound silicates.  This process enriches the surviving portions
of the dust grains in metallic iron, which further enhances the magnetically mediated collisions.
The magnetic enhancement to the collisional cross-section between the iron rich dust
 results in rapid grain growth leading to planetesimal formation.
While this process of knocking off silicates, which we term ``magnetic erosion'', occurs only in a very limited portion of
a protoplanetary disk, it is a possible explanation for Mercury's disproportionately large
iron core.
\end{abstract}


\section{Introduction}

It has long been known that Mercury is anomalously dense compared to the other rocky planets in our Solar System
\citep{Ash67, Howard74, Anderson87}.  Mercury must contain on the order of $70\%$ iron by mass for models
of its interior to match its density \citep{Lyttleton69, Harder01}, more than double the $\sim 30\%$ iron by mass
of the Earth and Venus \citep{Morgan80}.  Unsurprisingly, several models have been put forth to explain this oddity, covering
all the stages of planet formation, from the very condensation of the pre-Mercury solids \citep{Lewis72,EbelAlexander11} to
the stripping of silicates from an initially more Earth-like young Mercury.  Such stripping could be due to
the evaporation of Mercury's silicate surface in a hot Solar nebula \citep{Cameron85} or, perhaps
more crudely, it could be the consequence of a giant impact \citep{Benz88,Stewart13}.

The \Mes mission's recent measurements of Mercury's K/Th ratio has ruled out many models for Mercury's iron content however, so in this paper we introduce a new model
that acts during the very first stages of planet formation, when tiny dust grains collide and stick.
By considering the implications of metallic iron's electrical conductivity and, crucially,
its ferromagnetism, we find that ambient magnetic fields induce magnetic dipole moments in dust grains
rich in metallic iron which  lead to those dust grains colliding preferentially and violently.  The violence of the collisions
can knock silicates off the metallic iron rich grains in a process we name ``magnetic erosion'', and if so,
any planetesimal formed from the surviving grains will be enriched in iron.  The ambient magnetic field requirement
combines with iron's Curie temperature to strongly limit the orbital positions where magnetic erosion can act. This
is appropriate: of the rocky planets, only Mercury is so iron rich.

\section{Existing models for Mercury}

\subsection{Conventional models}

The \Mes \emph{GRS} instrument found terrestrial K/Th ratios on the surface of Mercury \citep{Peplowski11}.
Thorium has a $50\%$ condensation temperature $\sim 1660$K, significantly higher than the $\sim 1000$K
and $\sim 1350$K of potassium and silicon respectively \citep{Lodders03}.  Any model for Mercury's abundances
that relies on condensation or evaporation of silicates would predict a very low K/Th ratio because the potassium is more volatile
than silicon: evaporating silicon means also evaporating potassium.  Thus, the \Mes
measurements contradict volatility based models for Mercury's high iron/silicate mass ratio.

This rules out most of the conventional models excepting perhaps the giant impact model.
As discussed in \cite{Stewart13}, even though a giant impact would remove a significant amount of
the proto-Mercury's surface material, that (volatile depleted) material need not have been  reaccreted onto Mercury.
Note that the volatile depleted material that became the Moon did not reaccrete onto the Earth.
In such a scenario, if the depleted material was not reaccreted, then Mercury's K/Th ratio would not have been affected.
One should however consider that the scenario requires about a full Mercury's mass worth of depleted material.
That posited material is clearly not present today; and it is not obvious how to remove that much mass if it,
like Earth's Moon, remained gravitationally bound.

\subsection{Photophoresis}

While the giant impact model remains a promising candidate for explaining Mercury's iron content, it is also possible
that the iron enrichment occurred far earlier in Mercury's formation history, when tiny dust grains were first assembled
into larger boulders and planetesimals.
A recently introduced model that acts during that early formation stage relies on a key difference between iron metal and silicates: iron conducts heat far
more rapidly.  This causes photophoresis to treat the two differently \citep{Wurm13}.  
Photophoresis relies on the creation and maintenance of a temperature
gradient through dust grains and that work suggests that the higher thermal conductivity
of iron metal erased any temperature gradient in metallic iron rich grains in the Solar Nebula.  While
photophoresis transported silicate grains outwards, the metallic iron rich grains stayed behind, forming Mercury.

However, photophoresis relies on dust grains not rotating rapidly, which would smooth out the dust's internal
temperature gradient (but see \citealt{Eymeren}).  Further, photophoresis could only process dust in optically thin
regions of the Solar Nebula which saw sunlight
\citep{Wurm13}.  The high opacity of gas-dust mixtures places strong limits on where photophoresis can
act: for example photophoresis can operate near the dust sublimation radius because the opacity of the dust
free gas interior to that position is low;
and it can operate in the surface layers of a disk where the column densites to the star are low.  Surface layers however
are thin and contain only a small fraction of the vertically integrated solid material, while the sublimation radius is at a specific location
(albeit one which should vary with time).

Photophoretic sorting is a tentative but exciting proposal for explaining Mercury's iron content that does not contradict the \Mes
observations.  It also suggests we should also consider other differences between iron rich and
iron poor grains beyond their thermal conductivity, linking the outcome of planet formation back to
the first stages of dust growth.  In this paper, we consider the electrical conductivity and ferromagnetism of iron
rich dust.

\section{Overview}
\label{Basic}

\subsection{Dust dynamics}
\label{Sec_Dust_Dynamics}

The first stage of planet formation (in the core accretion scenario) is collisional coagulation of small
dust grains \citep{Blum08}.  During
this stage, dust grains
collide and stick, growing from sub-micron initial sizes to decimeter sizes or larger.  Once they grow that large,
collective processes such as the streaming instability \citep{Johansen07} can establish themselves, leading to a rapid concentration
of the dust and subsequent gravitational collapse of the dust concentrations.

The collisional growth phase is generally
taken to have two sub-phases: very small (up to several tens of microns across) dust grains collide due to Brownian motion, while larger dust grains cease
to couple perfectly to the disk turbulence, and so collide with each-other because of that turbulence
\citep{Volk80}.  The turbulence induced dust-dust collisions are much faster than those due to Brownian motion, and the turbulence
induced dust-dust collision speed increases as dust grains grow in size: larger dust grains can partially slip
through larger scale, faster moving turbulent structures.
We will refer to dust grains small enough that they collide primarily due to Brownian motion
as ``microphysical'', while ``macrophysical'' dust grains collide due to turbulence.

Protoplanetary disks are generally assumed to be turbulent, because whatever drives the accretion flow is either
turbulence itself, or will also drive turbulence.  In the standard $\alpha$ formalism \citep{SS73}, the
turbulent viscosity is parameterized as $\nu_t = \alpha c_s H$ where $H$ is the local gas scale height.  At $0.387$ A.U. from the
proto-Sun, assuming a minimum mass solar nebula (MMSN, \citealt{Hayashi85}) and $\alpha=10^{-4}$, this
implies that the smallest, dissipation scale turbulence had time and velocity scales of $t_{\eta}\simeq 73$\tp s and 
$v_{\eta} \simeq 27$\tp cm/s respectively.

Dust grains moving through gas feel a drag force, so that
\EQ
\frac{\partial \vv}{\partial t} = - \frac{\vv-\uu}{\tau},
\EN
where $\vv$ and $\uu$ are the dust and gas velocities respectively and $\tau$ is a friction timescale that lengthens
as the dust gets larger and more massive.
While the effect of turbulence on collisions between
very small grains (i.e.~grains with frictional stopping times $\tau$ shorter than the smallest turbulent timescale
$t_{\eta}$)
is as yet unclear \citep{Pan13}, dust grains of radius $a \sim 0.06$\tp cm and density $\rho=3$\op g/cm$^3$
have a frictional stopping time $\tau \sim t_{\eta}$.  This means that they are macroscopic
in our terms, and will collide with much smaller grains with
a turbulently induced collisional speed similar to that of the smallest scale turbulence, $v_{\eta} \sim 27$\tp cm/s, and with each
other at significantly lower, but still macroscopic, speeds \citep{Hubbard12, Hubbard13}.
We will therefore use $a=0.06$\tp cm as an order-of-magnitude lower bound on the size of the larger participant in a turbulently
induced dust-dust collision.

\subsection{Disk parameters}

The currently preferred
driver for observed accretion flows through protoplanetary disks is the Magneto-Rotational Instability (MRI, 
\citealt{Velikhov59,BH91}), which requires a minimum ionization fraction on the order of $10^{-13}-10^{-12}$ \citep{Gammie96}.
While protoplanetary disks are mostly neutral, they are nevertheless weakly ionized and can, in at least part of their volume,
support the MRI.
This ionization fraction can be met and surpassed in the inner disk at fractions of an A.U. where the temperature is above $T=1000$\op K (thermal ionization of
alkali metals), in the surface layers of the disk (UV and Xray photo-ionization) and in the outer disk, beyond $10-30$ A.U. 
(cosmic ray ionization).  The ionization also plays a role in dust collisions as will be discussed in \Sec{Sec-Charging}.

Where the MRI is active, it
naturally amplifies magnetic fields to plasma-$\beta$ values near $10$, depending on the height above the midplane \citep{Flock11}.
That $\beta$ value means that the magnetic fields have an energy density one tenth of the gas' thermal energy density.
Extrapolating from a MMSN
surface density profile, and assuming a background temperature of $1000$\op K,
$\beta=10$ implies a magnetic field\footnote{
Accretion disks have too few solids for their magnetization to play a dynamical role
on disk scales, so the $B$ and $H$ fields are interchangeable.  The $B$ field is standard but
 we are considering magnetized solids so the ambient field is more correctly reported as
$H=33$\op Oe.
}
 of $B=33$\op G for a surface density of $7000$ g/cm$^{-2}$, appropriate
at Mercury's orbit.  
This is more than $10^{-3}$ times the saturation field
for iron \citep{Brown58} at room temperature, which is itself
more than the saturation field for iron at $1000$\op K.  The relative permeability of iron at that temperature
range is several thousand \citep{Brown58}, so by multiplying those values together,
we see that the ambient field is adequate to
magnetically saturate iron.  This will become
important in \Sec{Magnetic}, but means that iron grains are expected to interact magnetically
even when the grains have many magnetic domains.

\section{Charging barrier}
\label{Sec-Charging}

Regardless of the source of the ionization, the thermal speed of electrons is higher than that of ions due to their vastly lighter mass.
As a result, all else being equal, the electrons collide with dust
grains at a faster rate, so the dust grains accumulate a net negative charge and repel each other.
This has strong implications for collisional growth of dust grains \citep{Simpson79,
Okuzumi09} because Brownian motion between dust grains is generally inadequate to overcome the electrostatic potential barrier
between negatively charged dust grains in regions of the disk with even modest ionization rates.
However, if macroscopic dust grains can be grown in the face of the charging barrier, then subsequent
macroscopically induced collisional velocities are high enough to overcome this barrier.

A possible way around this barrier uses the electrical dipole moment that charged grains can induce in each other
\citep{Konopka05}, which requires electrically conductive grains.  Note that in what follows we assume compact, as opposed to fluffy,
 grains. The charging barrier is much more significant for low-density grains, as they have a larger potential
barrier and shorter frictional stopping times than compact grains of the same mass  \citep{Okuzumi11}.

\subsection{Brownian motion}

Even while dust grains accrete a negative net charge, they also act as sites for recombination between the electrons and ions,
and as a result the
details of the grain charging are highly non-trivial if the ionization rate is low \citep{Okuzumi11, Matthews12, Ilgner12}.
Those details are,
however, beyond the scope of this paper.

Accordingly, we assume that there are enough free electrons in the gas that charging dust grains doesn't significantly affect the
supply.  In this case, the number densities of electrons and positively charged ions in the gas are approximately equal,
and the thermal velocity of those electrons is over $30$ times faster than that of the ions.  The steady state charge on the dust
is achieved when the dust-ion collision rate equals the dust-electron collision rate.  Given the very large velocity difference
between electrons and ions, and their equal number density,
this requires reducing the dust-electron collisional cross-section to near zero.
So, to within a factor of order unity
 the net charge on the dust
grains will be such that the Coulomb barrier faced by an electron approximately balances the electron's thermal energy:
\EQ
\frac{e q(a)}{a} \sim k_B T, \label{q(a)}
\EN
where $e$ is the elementary charge, $a$ is the radius of the grain, assumed spherical, and $q(a)$ is the net charge accumulated by the dust grain
\citep{Simpson79}.  If the ionization rate is high, and the supply of dust grains limited, $q(a)$ will be a factor of a few higher than suggested by \Eq{q(a)}, while in the opposite limit, $q(a)$ will be strongly suppressed \citep{Okuzumi11}.
It follows that in high electron availability regions,
 the potential barrier that must be overcome for two grains of radius $a_1$ and $a_2$ to collide is on the order of
\EQ
E_{Pot} \sim \frac{a_1 a_2}{a_1+a_2} \left(\frac{k_B T}{e}\right)^2. \label{Pot-charging}
\EN
Equating that energy with the thermal energy of the dust grains, $k_B T$, we finally find
\EQ
\frac{a_1 a_2}{a_1+a_2} \sim \frac{e^2}{k_B T} \sim 1.7 \times 10^{-2} \left(\frac{1000\op \rm{K}}{T}\right) \mu m \label{a-limit-charging}
\EN
to be the upper limit of dust grain sizes for which thermal motion allows collisional growth in the presence of grain charging
in regions with significant ionization.

If such a ceiling on dust size held, it would halt dust growth and subsequent planet formation
processes.  This size is also below the expected interstellar value of a fraction of a micron \citep{Draine03}, so unless the initial grain supply
is already large enough to experience turbulence induced dust-dust collisions, or other physics such as magnetic interactions
applies, collisional dust growth is not expected in highly ionized regions.

Metal rich grains can rearrange their charge, so pairs of charged metal grains will induce electrical dipole moments which attract each other,
weakening this charging barrier to grain growth.  In the limit of equally sized/equally
charged spheres, this reduces the potential barrier by $\sim 40\%$ although dissimilar spheres with the same
sign charge can even attract each other \citep{charged-spheres, Lekner12}.  However,
from \Eq{a-limit-charging} it is clear that the charging barrier is simply too high for such an effect to allow for collisional
growth for micron-sized grains.  Indeed, the charging barrier will prevent such grains from approaching much closer
than $6-60 \mu m$ ($T=100-1000\tp \rm{K}$), much larger than their own size and
so too far for induced dipoles to be dominant. 
It follows that the electrical conductivity
of grains with significant metal veins will not play a major role.

\subsection{Macroscopic motion}
\label{Charging-macro}

To overcome the charging barrier, two grains colliding at speed $v$ need to satisfy
\EQ
\frac 12 \frac{m_1 m_2}{m_1+m_2} v^2 > \frac{a_1 a_2}{a_1+a_2} \left(\frac{k_B T}{e}\right)^2, \label{charging-macro-cond}
\EN
where $m_1, m_2$ are the masses of the two grains.  
If $a_1>0.06$\tp cm$\tp \gg a_2$ then the larger grain $a_1$
is big enough to couple collisionally to turbulence, and $v > 27$\tp cm/s is a reasonable collision speed 
(see \Sec{Sec_Dust_Dynamics}). In that case, \Eq{charging-macro-cond} reduces to
\EQ
a_2>  \left(\frac{\rho}{3\op \rm{g/cm}^3}\right)^{-1/2} \frac{T}{1000\op \rm{K}}  \tp 0.04 \tp \mu m.
\EN
This is still larger than the limit set by \Eq{a-limit-charging}, but well satisfied by interstellar dust.
Further the condition set by \Eq{charging-macro-cond} is trivially satisfied for equal-size turbulently induced collisions.
As a consequence, we can see that even if iron-rich grains could preferentially bypass the charging barrier due to their
magnetization, the charging barrier becomes irrelevant as soon as
turbulently induced collisions become important.  Any preferential metallic iron-rich grain
growth will be rapidly forgotten as microscopic silicate grains are swept up by the macroscopic iron-rich ones.  
Making iron rich dust grains requires not just preferential iron-iron grain interactions, but also removing
the silicates that will inevitably agglomerate.

\section{Magnetic Erosion}
\label{Magnetic}

\subsection{Magnetically mediated collision speeds}
\label{MMCS}

Magnetic interactions between magnetized iron grains have been known to lead to rapid coaggulation since
 \citet[with corrections in \citealt{Nuth95}]{Nuth94}.  Indeed, the magnetic dipole-dipole interactions can enhance collisional cross-sections
by orders of magnitude, as confirmed by further experimental and numerical work \citep{Nubold02, Dominik02}.
Work on the growth of dust grains has focused on single-domain sized grains because saturated
magnetization is guaranteed in that case \citep{Wang10}.
It should however be noted that the ambient magnetic field can be quite significant
(\Sec{Basic}, see also \citealt{Fu12}), so induced
magnetic dipole moments can become important. 

In this section we approximate dust grains as compact spheres as opposed to fractal structures.  We are
in a regime where the charging barrier is important, so large fractal silicate structures are not expected.
Iron rich structures on the other hand are strongly magnetically bound, and collide with each other at speeds
above fragmentation speeds of silicates, and so also above their compaction speed.  We expect this
to also lead to compaction of iron structures, but that remains to be confirmed by experiment.

We leave the question of the strength of the magnetization for \Sec{caveats} and calculate the collisional velocity
that results from the collision of two magnetized grains.  We take our grains to be composed of iron metal (subscript ``i'')
 and ``other'' material (i.e.~silicates, subscript ``o''), with iron having density $\rho_i=7.86$ g/cm$^3$ and the other material $\rho_o$ with
$\rho_i/\rho_o=8/3$.  The two dust grains 
are taken to be spherical and have volume $\forall_{1,2}=(4\pi/3) a_{1,2}^3$.  The metallic iron components themselves are also taken to
be spherical, with volume $\forall_{i1,2}=(4\pi/3) r_{1,2}^3$.  The mass of the metallic iron component of a grain is parameterized as
$f \equiv m_i/m$ where $m$ is the total grain mass, and $f$ is identical for the two particles for convenience.
It follows that we can write the volume of a grain in terms of $m$ and $f$:
\EQ
\forall=\frac{m}{\rho_i} \left[f+(1-f)\frac{\rho_i}{\rho_o}\right].
\EN
We can further calculate
\EQ
\frac{\forall_i}{\forall} = \frac{f \phantom{\left[\frac{\rho_i}{\rho_0}\right]}}{\frac{\rho_i}{\rho_0}-\left[\frac{\rho_i}{\rho_0}-1\right]f}. 
\label{vol_rat}
\EN
Under our approximations, $f=0.1$ implies a metallic iron volume fraction of $4\%$, easily achieved by
chondritic meteorites \citep{Weisberg06}, and therefore a quite conservative estimate for small metallic iron
rich grains in the Solar Nebula.  Indeed, $f=0.2$ only requires a metallic iron volume fraction of $8.6\%$.

We assume optimal alignment between the magnetic dipole moments of two colliding dust
grains: the moments are parallel to each other and to the separation vector between the iron sub-grains.
While dust grains in protoplanetary disks rotate, the ambient magnetic field will align the grains, so rotation of the
dust grains will also be aligned and won't affect the magnetic interactions.  As we will show, the magnetic
interactions lead to ``large'' ($\sim 100$ cm/s) collision velocities, so any residual rotation not parallel
to the ambient field is unlikely to be fast enough to affect the outcome: the final approach will occur on a time scale
of a few dust grain radii divided by the collision speed.

Assuming this optimal alignment between the dipole moments, 
at the moment of impact
the velocity associated with the magnetic dipole interactions is given by
\EQ
\frac 12 \frac{m_1 m_2}{m_1+m_2} v^2=2 \frac{\forall_{i1} \forall_{i2} Z^2 M^2}{d^3}, \label{v-base}
\EN
where $M=1720$ emu/cc$^3$ \citep{Brown58} is the maximum magnetization of iron at room temperature
and $Z$ is the fractional magnetization of the dust grains compared to $M$,
assumed identical for the two particles for convenience.  In \Eq{v-base}, $d$ is the distance between the centers of the
two iron sub-grains, which need not be at the center of the dust grains.
Note that dipole interaction energies fall off as the cube of distance, so nearly all of the magnetic potential energy is released
in the last few dust-grain radii of the approach.  Dust grains have a solid density some $10^8$ time larger than the assumed
gas density at Mercury's position ($\sim 3 \times 10^{-8}$ g/cm$^3$), so while interactions with gas may bring grains close enough
together to interact, gas drag can safely be neglected in \Eq{v-base}.

We consider a total of four geometries deriving from two sets of limits. 
We consider the location of the iron
sub-grains, putting them at the center of the dust grains and on the surface of the dust grains at the location of impact.
The later case provides an upper limit to the magnetically induced collision speed,
while the first one captures both compact iron cores which have
accreted silicate rims, and homogenous iron metal/silicate mixtures.
Restricting ourselves to these cases simplifies the distance $d$ between
the centers of the two iron sub-grains: either $d=a_1+a_2$ (iron sub-grains at the center of their host) or $d=r_1+r_2$ (iron sub-grains
touching at impact).
We also consider the relative grain size:
identically sized grains compared with very different sized grains, i.e.~$r_1, a_1 \gg r_2, a_2$.

Under those conditions, we find
\EQ
\frac{v}{v_f}=\begin{cases}
         13 \frac{fZ}{\left([8/3]-[5/3]f\right)^{1/2}}&  
               \text{same/center,} \\[0.25 cm]
        13 \sqrt{f} Z  & 
               \text{same/impact,} \\[0.2 cm]
         25 \frac{fZ}{\left([8/3]-[5/3]f\right)^{1/2}}  &  
               \text{different/center,} \\[0.25 cm]         
         25 \sqrt{f} Z  &
               \text{different/impact,} \\
     \end{cases}
     \label{collision_formula}
\EN
where $v_f=100$\tp cm/s is the estimated collisional velocity for fragmentation of loosely bound silicate grain
agglomerations \citep{Guttler10}.
Interestingly the sizes of the particles cancel out completely under our choice of parameterization.
For $f=0.1$ and $Z=1$ this reduces to
\EQ
\frac{v}{v_f}=\begin{cases}
  0.79   &               \text{same/center,} \\
   3.97 &               \text{same/impact,} \\
   1.58 &               \text{different/center,} \\      
    7.94&               \text{different/impact.} \\
    \end{cases}
\EN
All these cases have an adequate velocity to satisfy
\Eq{charging-macro-cond} for grains we consider, so the magnetic interaction overpowers the charging barrier.

Crucially, except for identical particles with iron at the center, these
values rise well above the critical velocity for destruction of loosely bound grains, $v_f=100$ \tp cm/s.
Indeed, for $f=0.1$, values of $Z>0.125$ can lead to destructive collisions, albeit only just, and only in
otherwise optimal conditions. 

\subsection{Competition between charging and magnetic interactions}

Magnetic interactions are strong, leading to macroscopic collisional velocities, which were shown in \Sec{Charging-macro}
to be able to overcome the charging barrier.  However, magnetic dipole interaction energy falls
off with the cube of distance, while the electrostatic energy barrier falls of as the inverse distance.  If the magnetic energy
still dominates at the distance when
the electrostatic energy barrier between two grains is equal to the thermal energy, then we can be confident
that the charging barrier does not prevent magnetic erosion.  We can calculate this distance $d$ for identical grains
of radius $a$ using \Eq{Pot-charging}:
\EQ
d = \frac{k_B T}{e^2} a^2. \label{charging_distance_limit}
\EN
Setting the left hand side of \Eq{v-base} to the thermal energy $k_B T$ and using \Eq{charging_distance_limit}
we find
\EQ
Z \left(\frac ra\right)^3 = Z \frac{\forall_i}{\forall} = \frac{3 k_B^2 T^2}{\sqrt{2} 4 \pi M e^3} \simeq \frac{1}{60}.
\EN
This is generally fulfilled, and our definition of $d$ did not consider the attractive magnetic interaction so it is also
a slight overestimate of the effect of the charging barrier.  Nevertheless, the charging barrier can effect
the rate at which microscopic,
low iron content grains interact.  For example, for $f=0.1$ we have $\forall_i/\forall=0.04$ so to fulfill this criterion
we need $Z=0.4$; and somewhat above-average thermal energy grains might be needed to
overcome the charging barrier.

\begin{figure}[t!]\begin{center}
\includegraphics[width=\columnwidth]{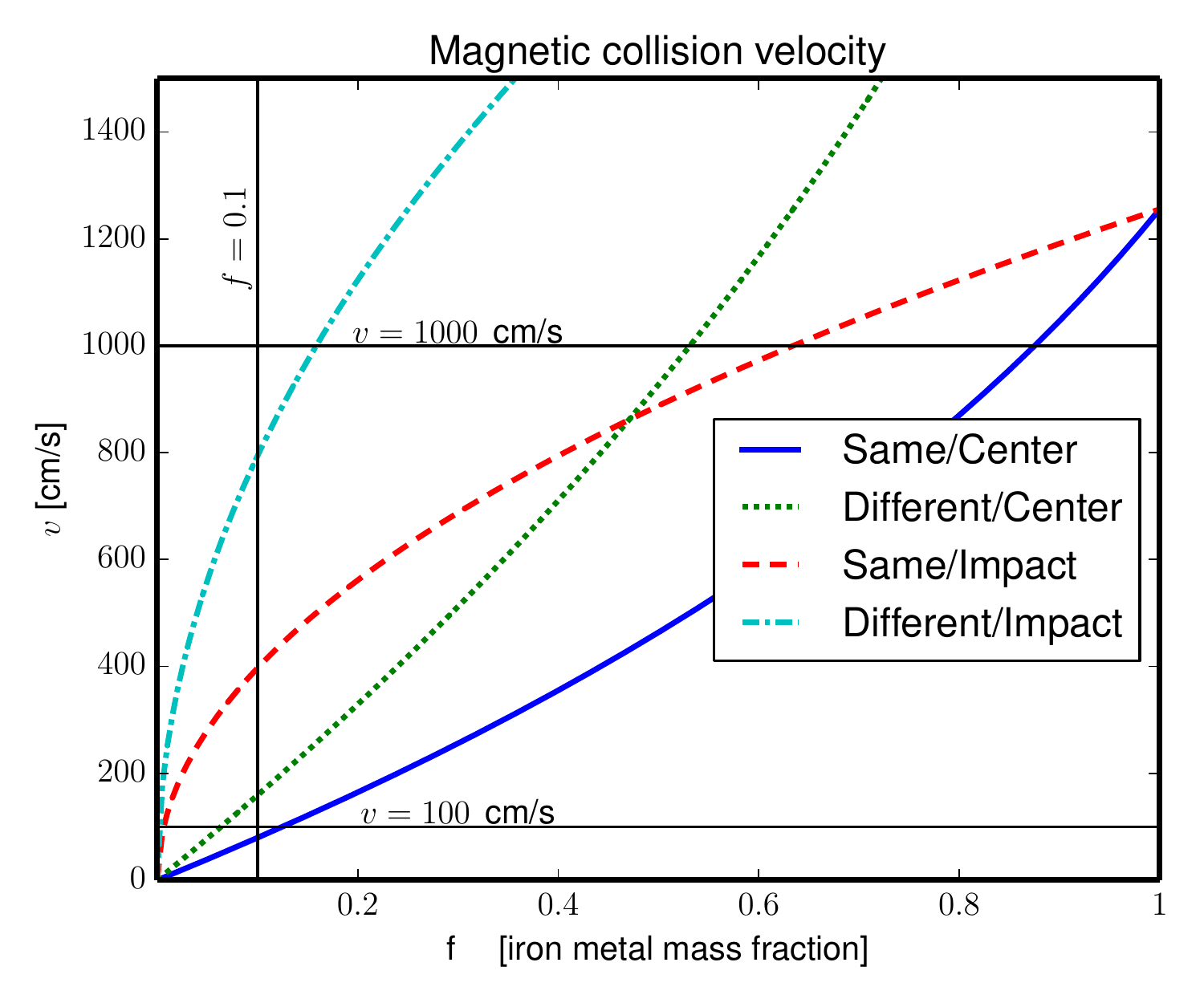}
\end{center}\caption{
Plots of the magnetically induced collision velocities from \Eq{collision_formula} as a function of the metallic iron
mass fraction $f$, for $Z=1$.  Thin lines are $f=0.1$, $v=100$\op cm/s and $v=1000$\op cm/s.  Recall that
$v\propto Z$.
\label{velocities} }
\end{figure}

\subsection{Effect of collisions} 

These destructive collisions will not knock apart the iron grains, which are
bound together magnetically with precisely the energy required to generate the collision speed $v$.
However, if $v>v_f$, the collisions 
could be violent enough to knock off silicate grains attached to the agglomeration
through mere Van der Waals forces (which is the source of the estimates for $v_f$).
It immediately follows that induced magnetic dipole-dipole interactions are a way to
cause metallic iron-rich grains to collide preferentially, while simultaneously removing
silicates.
Once the iron rich grains become macroscopic, they can bypass the charging barrier
and accrete silicate rims, which need to be eroded in subsequent collisions
as noted in \Sec{Charging-macro}.

\begin{figure}[t!]\begin{center}
\includegraphics[width=\columnwidth]{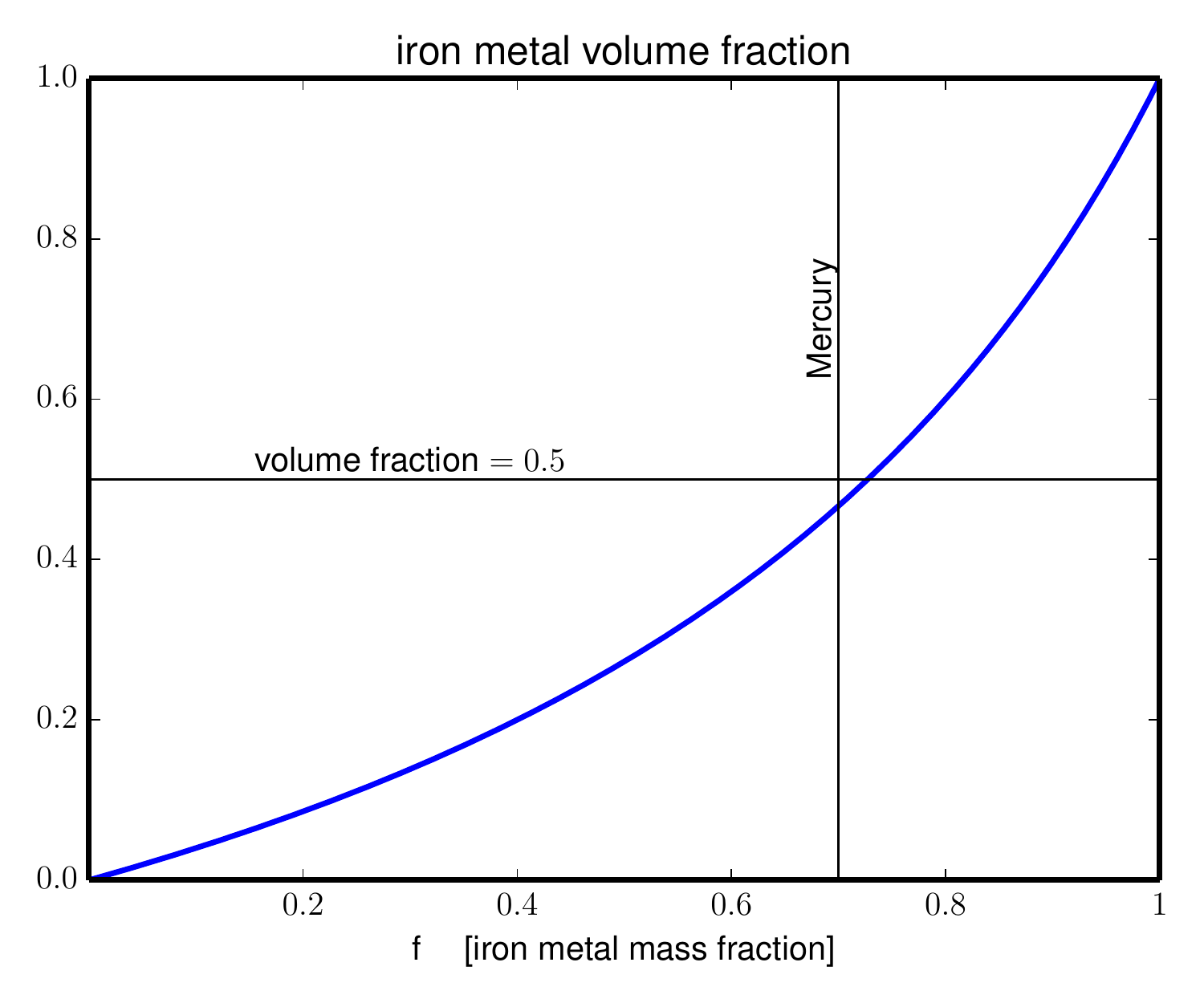}
\end{center}\caption{
Plot of the iron metal volume ratio as a function of the iron metal mass ratio ($\forall_i/\forall$).  Thin lines are
$\forall_i/\forall=0.5$ and Mercury's value of $f=0.7$.
\label{fig_vol_rat} }
\end{figure}

In the ``impact'' case, and for iron grains with silicate rims (one possible interpretation of the ``center'' case),
the silicates are loosely bound to the iron and free to be knocked off,
although the precise extent of the erosion will need to be studied both numerically and experimentally.
However, if the dust is made of a homogenous mixture of iron and silicate grains, then the iron (magnetically
bound) could trap silicates, allowing only a limited amount of erosion
from the surface.  In \Fig{fig_vol_rat} we plot the iron volume ratio from \Eq{vol_rat}.
This makes clear that at small $f$, iron makes up only a small fraction of the volume.  The iron
grains are unlikely to be able to do
much trapping until the volume ratio is near $50\%$, which happens if $f$ is reasonably close to the 
$0.7$ value of Mercury.

\subsection{Subsequent survival}

Turbulence clusters dust grains, but dust-dust intra-cluster velocities are low enough that there are few if any
intra-cluster collisions \citep{Hubbard12,Hubbard13}.  In the absence of non-turbulent effects, dust collisions
are controlled by rarer higher-velocity encounters.  However, the same magnetic interactions that lead
to the enhanced collision speeds
seen in \Fig{velocities} would cause the turbulently created (but otherwise collisionless) clusters to undergo coagulating
collisions with large effective cross-sections \citep{Nuth95}. 

While metallic iron rich dust grains can coagulate rapidly, they still need to grow very large to survive: in protoplanetary disks,
small grains are entrained in the accretion flow onto the protostar, while larger grains experience the
``meter barrier'' \citep{Weidenschilling77} and drift radially inwards even faster than the gas accretes.

We can check the ability of the grains to grow adequately large.  The magnetic collisional cross-section
between equal sized iron grains is larger than the geometric one by a factor of
\EQ
\frac{ \sigma_m}{\sigma} = 1+\frac{4 \pi}{3} \frac{Z^2 M^2}{\rho_i v^2}, \label{cross-sec}
\EN
where $v$ is the background (non-magnetic) velocity dispersion of the dust.  To estimate the background velocity
dispersion $v$ at which magnetic interactions cease to significantly enhance the dust growth rate, we solve \Eq{cross-sec} for
$\sigma_m=2\sigma$, assuming $Z=0.5$, and find $v=600$ cm/s.  This is comparable to both the dust velocity
dispersion caused by turbulence (assuming $c_s \simeq 2 \times 10^5$ cm/s and $\alpha=10^{-4}$)
 and to the velocity dispersion caused by headwind induced drift for particles with frictional drag time 
$\tau \sim 0.1/\Omega_K$, where $\Omega_K$ is the local Keplerian frequency.

Dust grains with that long
a frictional drag time $\tau$ are settled to a scale height $0.05 H_g$ where $H_g$ is the local isothermal gas
scale height.  This means that the iron dust-gas mass ratio near the midplane will be approximately $0.1$ (assuming
$30\%$ of the potential solids are iron, as witnessed by Earth), and the streaming instability can be triggered
 \citep{Johansen07}.  The streaming instability in turn further enhances the local metallic iron rich dust density to
 the point of gravitational instability, forming iron rich planetesimals which are large enough to survive
 the meter barrier.

\section{Caveats: Curie temperature, background fields and volatile abundances}
 \label{caveats}
 
Any model for Mercury's iron fraction faces a fine-tuning problem: there is only one Mercury, and it is small.
Thus, its iron enrichment mechanism must not be universal.  If the explanation for the iron core is the collisional
erosion of silicates from metallic iron rich dust grains due to magnetic dipole interactions, then such interactions
should be spatially restricted in a protoplanetary disk, or additional physics must be added to explain
the low iron content of, for example, the Earth relative to Mercury.
 
Clearly, if metallic iron rich grains are significantly magnetized, the resulting dipole-dipole interactions
have strong implications for the evolution of small particles.  However, magnetizing grains significantly
larger than a single magnetic domain requires strong external magnetic fields.  As we saw in \Sec{Basic},
strong ambient magnetic fields are expected in MRI active regions, particularly in the inner active region,
with comparatively high thermal energy densities.  It is, however, not obvious that those fields can be used
to magnetize iron-rich dust grains because the ambient temperature may be too high.

\subsection{Inner active region}

The Curie temperature of iron is $1043$\op K \citep{Brown58}, and thermal ionization of alkali metals
is only expected to allow the MRI to operate
above about $1000$\op K \citep{Gammie96}.  Gammie's calculation relies on thermal ionization of potassium
in the gas phase, but potassium has a $50\%$ condensation temperature at $10^{-4}$\op bar of $1006$\op K \citep{Lodders03}.  That
pressure is appropriate for the Solar Nebula at Earth's position and $T \simeq 300$K.  The higher densities and temperatures 
at Mercury's orbit imply a pressure higher
by over an order of magnitude, raising the potassium condensation temperature  by $25-50$\op K (\citealt{Ebel06}, Plate 1).
It follows that there is only a narrow temperature window where the MRI is active at the same time as the temperature is below the Curie
temperature of iron.  

Nevertheless, the initial and maximum relative permeabilities of iron in the temperature range of 
$1000-1043$\op K are $2$ to $7$ thousand, while its saturated magnetic field strength is reduced by a factor
of $2$ to $3$ from the room temperature value.  So even for magnetic fields of only a few Gauss,
the iron will  be magnetized to near saturation ($Z \simeq 0.3 - 0.5$), and iron grains
can be magnetized by remnant fields just outside the MRI active region. We should also
remember that disk temperatures are hardly uniform in space: the gravitational energy released by the accretion flow
is dissipated, at least in MRI-active regions, in narrow current sheets.  These current sheets
are higher temperature/lower magnetic field
regions, with low temperature/high magnetic field
counterparts below iron's Curie temperature even in the MRI-active region.

 The dissipation of magnetic energy in thermally ionized regions displays surprising complications such as
the short-circuit instability \citep{HubbardSI12,McNally13} and is not yet well understood.  As such,
determining the precise scale of the fine tuning problem faced by the magnetic erosion model requires a detailed understanding
of the magnetic properties of the grains involved along with a detailed understanding of the ambient thermal and
magnetic fields, and is beyond the scope of this paper.

\subsection{Other MRI-active regions}

There are other, cooler MRI-active regions, such as the surface layers of the disk which are photo-ionized.  As they are surface layers
however, there is little solid material in them, and dust grains that grow even modestly will settle out of them towards the midplane.
As a result, we do not anticipate significant magnetic erosion outside of the outer edge of the inner, thermally ionized MRI-active region.
On the other hand, the location of that edge varies with time, both secularly as the disk accretes and on shorter
time scales as accretion events such as FU Orionis outbursts heat the disk \citep{Martin12,Martin13}.

Even if the temperature was hot ($\sim 1000$\op K) significantly farther from the proto-Sun Mercury's orbit of $0.387$\op AU,
the gas density would be lower.  A plasma $\beta=10$ together with a surface density of $1700$\op g/cm$^{-2}$ and
a temperature $T=1000$\op K at $R=1$\op AU
imply an ambient field of $2$\op G, dramatically reducing the ability of the field to magnetize iron grains.  Non-thermal ionization,
associated with lower temperatures, imply even lower magnetic fields yet.

\subsection{Silicates and potassium}

While magnetic erosion will create iron-rich macroscopic dust grains in regions of the disk where it operates, silicate growth
could proceed in parallel, albeit slower.  For the faster growth of iron rich grains to matter in the long run, the silicate material
must be lost from the disk.  This can easily be achieved if the silicate grains remain small, and hence well coupled to the gas accretion flow through the
disk which will carry them away onto the star.  As we discussed
in \Sec{Sec-Charging}, the charging barrier is difficult for microscopic dust grains to bypass in regions of
relatively high ionization, and the outer edge of the inner MRI-active zone is precisely a location where the
charging barrier is expected to act.
As long as magnetic erosion acts fast enough to prevent macroscopic silicate agglomerations from forming on macroscopic
iron rich dust grains, the silicates will have
remained in microscopic form,
locked to the gas accretion flow and will have been lost to the proto-Sun, allowing the proto-Mercury
body, initially formed from iron-rich components, to remain iron rich.

Further, the \Mes results \citep{Peplowski11} imply that Mercury is
not particularly depleted in potassium, and yet we are suggesting that Mercury's large iron core is due to a process
that can only occur at least very near to regions where potassium is found in the gas phase, as opposed to the solid phase.  
However, potassium evaporates and becomes thermally ionized at nearly the same temperature:
microphysical potassium depleted silicates cannot grow in regions where magnetic erosion is active because they cannot
overcome the collisional electrostatic potential barrier due to the electrons from ionized potassium in the gas phase.
As long as Mercury's silicates were not formed
in a magnetic erosion active region, there is no conflict with the \Mes data.  In this model, the material that became
 Mercury's silicate mantle was formed either radially
outside of the region where the proto-Mercury iron-rich bodies formed, or later in time, after the disk cooled.

\section{Conclusions}

We have shown how the outer edge of the inner, thermally ionized MRI-active region of protoplanetary disks
has the correct conditions to strongly 
magnetize metallic iron rich dust grains.  This magnetization leads to energetic magnetically mediated collisions which
erode any non-magnetic component of the participating dust, leading to large, very iron-rich grains, appropriate for
explaining Mercury's anomalously high iron content.  The process can begin with even modest metallic iron fractions and
 results in extremely rapid dust growth due to the magnetically enhanced
collisional cross-sections, leading to planetesimal formation.

The model we have proposed is basic: we compared the magnetically mediated collision velocity to a critical threshold velocity derived
for other purposes.  While the critical velocity we use should be appropriate for iron cores with silicate rims, it
may overestimate the effect of magnetic erosion for homogenous mixtures of iron grains and silicates.  Future experimental
work on collisions that include ferromagnetic particles and possibly external magnetic fields will be needed
to move beyond our more order-of-magnitude results.

This process is unlikely to have taken place in other regions of the proto-Solar nebula because of the magnetic field requirement, and so does not conflict
with the evidence presented by the other rocky planets.  The process also does not conflict with (nor indeed interact
with at all) the
volatile abundances measured by \emph{MESSENGER}, unlike most extant models for Mercury's formation.  Furthermore,
while the photophoretic
model can only select for iron-rich grains, not create or enrich them,
 the magnetic erosion model allows for iron enrichment of grains, making Mercury possible even if the initial dust supply
 did not contain extremely metal-rich grains.

Even if magnetic fields likely did not play a major role in colliding macroscopic dust grains except in 
specific parts of
the Solar Nebula, they could have played a role for microscopic iron grains with small numbers of magnetic domains
as first discussed in \cite{Nuth94}.  Sub-micron interstellar iron grains with very finite numbers of magnetic domains
might have been concentrated into iron nodules because even disordered jumbles of magnetic domains
have residual magnetic dipole moments; and even quite weak magnetic interactions can generate far stronger
interaction speeds than Brownian motion \citep{Nuth95}.  There are indications of such processes in the meteoritical record:
a significant fraction of the variation in chondrule composition, even inside specific chondrites, is due to the varying
amount of iron \citep{Grossman82}.

\acknowledgements

This research has made use of the National Aeronautics and Space AdministrationÕs Astrophysics Data System Bibliographic Services.
The work was supported by National Science Foundation, Cyberenabled Discovery Initiative grant AST08-35734, 
AAG grant AST10-09802,
and a Kalbfleisch Fellowship from the American Museum of Natural History.
I thank Denton S. Ebel for his support, critiques and and helpful comments.


\begin{thebibliography}

\bibitem[Anderson et al.(1987)]{Anderson87} Anderson, J.~D., 
Colombo, G., Espsitio, P.~B., Lau, E.~L., 
\& Trager, G.~B.\ 1987, Icarus, 71, 337 

\bibitem[Ash et al.(1967)]{Ash67} Ash, M.~E., Shapiro, I.~I., 
\& Smith, W.~B.\ 1967, \aj, 72, 338 

\bibitem[Balbus 
\& Hawley(1991)]{BH91} Balbus, S.~A., \& Hawley, J.~F.\ 1991, \apj, 376, 214 

\bibitem[Benz et al.(1988)]{Benz88} Benz, W., Slattery, W.~L., 
\& Cameron, A.~G.~W.\ 1988, Icarus, 74, 516 

\bibitem[Blum 
\& Wurm(2008)]{Blum08} Blum, J., \& Wurm, G.\ 2008, \araa, 46, 2

\bibitem[Bozorth(1951)]{Bozorth51} Bozorth, R., Ferromagnetism,\ 1951, reprinted 1993
by IEEE Press, New York

\bibitem[Brown(1958)]{Brown58} Brown, W. F., Magnetic Materials,
Ch 8 in the Handbook of Chemistry and Physics, Condon and Odishaw, eds., McGraw-Hill, 1958

\bibitem[Cameron(1985)]{Cameron85} Cameron, A.~G.~W.\ 1985, 
Icarus, 64, 285 

\bibitem[Dominik \&
N{\"u}bold(2002)]{Dominik02} Dominik, C.,  N{\"u}bold, H.\ 2002, Icarus, 157, 173 

\bibitem[Draine 
\& Lazarian(1999)]{Draine99} Draine, B.~T., \& Lazarian, A.\ 1999, \apj, 512, 740 

\bibitem[Draine(2003)]{Draine03} Draine, B.~T.\ 2003, \araa, 41, 241 

\bibitem[Ebel(2006)]{Ebel06} Ebel, D.~S.\ 2006, Meteorites and 
the Early Solar System II, 253 

\bibitem[Ebel 
\& Alexander(2011)]{EbelAlexander11} Ebel, D.~S., \& Alexander, C.~M.~O.\ 2011, \planss, 59, 1888 

\bibitem[van Eymeren 
\& Wurm(2012)]{Eymeren} van Eymeren, J., \& Wurm, G.\ 2012, \mnras, 420, 183 

\bibitem[Flock et al.(2011)]{Flock11} Flock, M., Dzyurkevich, 
N., Klahr, H., Turner, N.~J., \& Henning, T.\ 2011, \apj, 735, 122 

\bibitem[Fu 
\& Weiss(2012)]{Fu12} Fu, R.~R., \& Weiss, B.~P.\ 2012, Journal of Geophysical Research (Planets), 117, 2003 

\bibitem[Gammie(1996)]{Gammie96} Gammie, C.~F.\ 1996, \apj, 457, 
355 

\bibitem[Grossman 
\& Wasson(1982)]{Grossman82} Grossman, J.~N., \& Wasson, J.~T.\ 1982, \gca, 46, 1081 

\bibitem[G{\"u}ttler et 
al.(2010)]{Guttler10} G{\"u}ttler, C., Blum, J., Zsom, A., Ormel, C.~W., \& Dullemond, C.~P.\ 2010, \aap, 513, A56 

\bibitem[Harder 
\& Schubert(2001)]{Harder01} Harder, H., \& Schubert, G.\ 2001, Icarus, 151, 118 

\bibitem[Hayashi et al.(1985)]{Hayashi85} Hayashi, C., Nakazawa, 
K., \& Nakagawa, Y.\ 1985, Protostars and Planets II, 1100 

\bibitem[Howard et al.(1974)]{Howard74} Howard, H.~T., Tyler, 
G.~L., Esposito, P.~B., et al.\ 1974, Science, 185, 179 

\bibitem[Hubbard(2012)]{Hubbard12} Hubbard, A.\ 2012, \mnras, 
426, 784 

\bibitem[Hubbard et al.(2012)]{HubbardSI12} Hubbard, A., McNally, 
C.~P., \& Mac Low, M.-M.\ 2012, \apj, 761, 58

\bibitem[Hubbard(2013)]{Hubbard13} Hubbard, A.\ 2013, \mnras, 
432, 1274 

\bibitem[Ilgner(2012)]{Ilgner12} Ilgner, M.\ 2012, \aap, 538, A124 

\bibitem[Johansen et al.(2007)]{Johansen07} Johansen, A., Oishi, 
J.~S., Mac Low, M.-M., et al.\ 2007, \nat, 448, 1022 

\bibitem[Konopka et al.(2005)]{Konopka05} Konopka, U., Mokler, 
F., Ivlev, A.~V., et al.\ 2005, New Journal of Physics, 7, 227

\bibitem[Lekner(2012)]{Lekner12} Lekner, J.\ 2012, Measurement 
Science and Technology, 23, 085007 

\bibitem[Lewis(1972)]{Lewis72} Lewis, J.~S.\ 1972, Earth and Planetary Science Letters, 15, 286 

\bibitem[Lodders(2003)]{Lodders03} Lodders, K.\ 2003, \apj, 591, 
1220 

\bibitem[Lyttleton(1969)]{Lyttleton69} Lyttleton, R.~A.\ 1969, \apss, 5, 18 

\bibitem[Martin 
\& Livio(2012)]{Martin12} Martin, R.~G., \& Livio, M.\ 2012, \mnras, 425, L6 

\bibitem[Martin 
\& Livio(2013)]{Martin13} Martin, R.~G., \& Livio, M.\ 2013, \mnras, 434, 633 

\bibitem[Matthews et al.(2012)]{Matthews12} Matthews, L.~S., Land, 
V., \& Hyde, T.~W.\ 2012, \apj, 744, 8 

\bibitem[McNally et al.(2013)]{McNally13} McNally, C.~P., 
Hubbard, A., Mac Low, M.-M., Ebel, D.~S., 
\& D'Alessio, P.\ 2013, \apjl, 767, L2 

\bibitem[Morgan 
\& Anders(1980)]{Morgan80} Morgan, J.~W., \& Anders, E.\ 1980, Proceedings of the National Academy of Science, 77, 6973 

\bibitem[N{\"u}bold et al.(2002)]{Nubold02} N{\"u}bold, H., 
Poppe, T., Dominik, C., 
\& Glassmeier, K.-H.\ 2002, Advances in Space Research, 29, 773 

\bibitem[Nuth et al.(1994)]{Nuth94} Nuth, J.~A., III, Faris, 
J., Wasilewski, P., \& Berg, O.\ 1994, Icarus, 107, 155 

\bibitem[Nuth 
\& Wilkinson(1995)]{Nuth95} Nuth, J.~A., III, \& Wilkinson, G.~M.\ 1995, Icarus, 117, 431 

\bibitem[Okuzumi(2009)]{Okuzumi09} Okuzumi, S.\ 2009, \apj, 698, 
1122 

\bibitem[Okuzumi et al.(2011)]{Okuzumi11} Okuzumi, S., Tanaka, 
H., Takeuchi, T., \& Sakagami, M.-a.\ 2011, \apj, 731, 95

\bibitem[Pan 
\& Padoan(2013)]{Pan13} Pan, L., \& Padoan, P.\ 2013, arXiv:1305.0307 

\bibitem[Peplowski et 
al.(2011)]{Peplowski11} Peplowski, P.~N., Blewett, D.~T., Denevi, B.~W., et al.\ 2011, \planss, 59, 1654

\bibitem[Righter et al.(2006)]{Righter06} Righter, K., Drake, 
M.~J., 
\& Scott, E.~R.~D.\ 2006, Meteorites and the Early Solar System II, 803 

\bibitem[Shakura 
\& Sunyaev(1973)]{SS73} Shakura, N.~I., \& Sunyaev, R.~A.\ 1973, \aap, 24, 337 

\bibitem[Simpson et 
al.(1979)]{Simpson79} Simpson, J.~C., Simons, S., \& Williams, I.~P.\ 1979, \apss, 61, 65 

\bibitem[Stewart et al.(2013)]{Stewart13} Stewart, S.~T., 
Leinhardt, Z.~M., 
\& Humayun, M.\ 2013, Lunar and Planetary Institute Science Conference Abstracts, 44, 2306 

\bibitem[Thomson, W. 1853]{charged-spheres} Thomson, W.\ 1853,\ 
On the mutual attraction or repulsion between two electrified spherical
conductors, pp. 86Ð97. In Reprint of papers on electrostatics and magnetism. London, UK:
Macmillan, 1884.

\bibitem[Velikhov(1959)]{Velikhov59} Velikhov, E. P. 1959, Soviet Phys.-JETP Lett., 36, 995

\bibitem[Voelk et 
al.(1980)]{Volk80} Voelk, H.~J., Jones, F.~C., Morfill, G.~E., \& Roeser, S.\ 1980, \aap, 85, 316 

\bibitem[Wang et al.(2010)]{Wang10} Wang, M., Chen, Q., 
\& Ding, Q.\ 2010, Journal of Geophysical Research (Planets), 115, 5005 

\bibitem[Weidenschilling(1977)]{Weidenschilling77} Weidenschilling, 
S.~J.\ 1977, \mnras, 180, 57 

\bibitem[Weisberg et al.(2006)]{Weisberg06} Weisberg, M.~K., 
McCoy, T.~J., 
\& Krot, A.~N.\ 2006, Meteorites and the Early Solar System II, 19 

\bibitem[Wurm et al.(2013)]{Wurm13} Wurm, G., Trieloff, M., 
\& Rauer, H.\ 2013, \apj, 769, 78 

\end{thebibliography}
\end{document}